Because of its inherent 2-D character, the eigenvalue equation for the hindered rotation around a normal cation site of an off-centered impurity nearest-neighboring an F center is the well-known Mathieu equation. We present an overview of literature data on Mathieu's periodic functions providing exact solutions to the $Li^+$ reorientational problem. We compare them with bottom-well approximating solutions by harmonic oscillator functions at an effective vibrational frequency renormalized by both first- and third- order electron-mode coupling. We finally discuss the in-plane inversion polarizability of an off-center impurity assumed to form a dipole-dipole coupling with a nearby F center.

1. Introduction

Following the general arguments of Reference [1], we now simplify the eigenvalue problem through freezing-in one of the vibrational coordinates: By setting $\theta = \frac{1}{2}\pi$, we consider the hindered rotation of an off-centered impurity in the equatorial plane. Thereby, we convert the original symmetry from octahedral $O_h$ ($T_{1u}$ symmetry-breaking vibrational mode), pertinent to an isolated $Li^+$ impurity, to axial $C_{4v}$ ($E_u$ symmetry-breaking mode), pertinent to a $Li^+$ impurity nearest-neighboring an F center. Physically, the formal $O_h$ to $C_{4v}$ symmetry lowering results from the presence of a nearby F center but it also leads to the immobilization of the vibrating halogen pair along the $Q_Z$-axis if the F center is seated in the [001] site [2]. The planar problem is much easier to deal with, the eigenvalue equation turning into the familiar Mathieu equation [3] whose solutions are well-tabulated. We also compare the exact solutions through Mathieu's periodic functions with harmonic-oscillator functions approximating for the rotational eigenstates near the bottom of the reorientational wells, as applied earlier to impurity tunneling [4]. However, the latter eigenstates are now controlled by an effective vibrational frequency renormalized by both first- and third- order electron-mode coupling.

As an off-centered $Li^+$ impurity is placed near an F center in ground electronic state, a dipole-dipole coupling arises, since both species are electrostatically polarizable. While the F center polarizability has been estimated using static electronic wave functions [2], the off-center $Li^+$ polarizability incorporated vibronic (inversion) counterparts. An improved expression for the vibronic polarizability is now discussed pertaining to the in-plane rotation with four reorientational off-center sites [5].

## 2. Off-axis impurity at $F_A$ center

### 2.1. Reorientational sites

According to the foregoing model theory an impurity ion goes off-center due to electronic states' mixing by a $T_{1u}$ vibrational mode of the three diametrical <100> halogen pairs nearest-neighboring the normal cation site [2]. Coupling to that same vibrational mode has also been seen to drive the impurity rotation and produce hindering potentials to control that rotation. With an F center substituting for one of these halogens as in an $F_A$ center, the impurity motion turns basically two-dimensional (2D) in the perpendicular plane, since the corresponding vibrating pair is demobilized. With an F center at [001] the impurity plane will be (100) $\theta = \frac{1}{2}\pi$ and the effective 2D rotational Hamiltonian turns out to be

$$H_{vib}(2D) = -(\eta^2/2I_A)(\partial^2/\partial\varphi^2) \pm (I_A\omega_A^2/b)\{(d_c - d_b)[\tfrac{1}{4}(3+\cos(4\varphi))] + d_b\}Q_A^2 +$$

$$\tfrac{1}{2}[(1\pm 2)(b^2/M_A\omega_A^2) - M_A\omega_A^2 E_{\alpha\beta}^2/4b^2] \qquad (1)$$

($\eta = h/2\pi$) introducing the moment of inertia $I_A = MQ_A^2$ of the rotating ion. It seems worth noting that similar reorientation-hindering potentials of restricted rotators have been assumed in the past [6,7]. In the above equation

$$Q_A = Q_0\{1 - (\alpha_F p_{sp}^2/E_{JT} k^2 R^6)/[(4E_{JT}/E_{\alpha\beta}^2 - 1]\} \qquad (2)$$

is the off-center radius of a $Li^+$ impurity at $F_A$ center which shrinks relative to the free Li impurity off-center radius as a result of the F center perturbation. $\alpha_F$ is the F center polarizability, $p_{sp}$ is the $a_{1g} - t_{1u}$ mixing dipole, k is an appropriate dielectric constant [2].

The first derivative of the $\varphi$-containing part of the potential in (1) is proportional to $-\sin(4\varphi)$ and the curvature to $-4\cos(4\varphi)$. It follows that there are eight extrema at $\varphi = n(\pi/4)$ with $n = 0, 1 \div 7$. We assume that $d_c - d_b < 0$. For the lower-sign (-) branch, four of the extrema are maxima at $n = 0,2,4,6$ and four are minima at $n = 1,3,5,7$. The minima are metastable rotational sites, while the maxima are barriers (saddle points) in-between. For the upper-sign (+) branch the extrema are virtually the same, though the minima and maxima interchange each other.

For brevity, we drop the suffix A in subsequent equations. The bottom of the lower-branch well is at energy

$$E_{Lmin} = -(I\omega_{renII}^2/8)[(d_b + d_c)/(d_b - d_c)] - E_{JT}[1 + (E_{\alpha\beta}/4E_{JT})^2] \qquad (3)$$

while the top of that well is at

$$E_{Lmax} = -(I\omega_{renII}^2/8)[d_c/(d_b - d_c)] - E_{JT}[1 + (E_{\alpha\beta}/4E_{JT})^2] \qquad (4)$$

For the upper branch the well bottom is at energy

$$E_{Umin} = (I\omega_{renII}^2/4)[d_c/(d_b - d_c)] + E_{JT}[3 - (E_{\alpha\beta}/4E_{JT})^2] \qquad (5)$$

and the top is at

$$E_{Umax} = (I\omega_{renII}^2 / 8) [(d_b + d_c)/(d_b - d_c)] + E_{JT}[3 - (E_{\alpha\beta}/4E_{JT})^2] \qquad (6)$$

The energy splitting of the two branches, i.e. the gap between a minimum on the upper branch and a maximum on the lower branch, is

$$E_{12} \equiv E_{Umin} - E_{Lmax} = 2\{(I\omega_{renII}^2 / 4)[d_c/(d_b - d_c)] + 2E_{JT}\} \qquad (7)$$

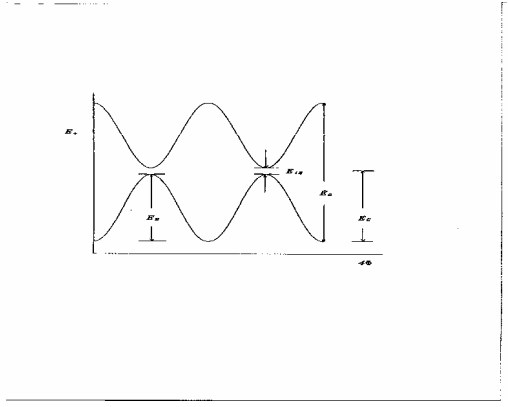

Figure 1. Azimuthal dependence (illustrative) of the dual-branched vibronic potential energy entering into equation (1). Various rotational quantities are indicated as introduced in text.

This gap controls the reorientational rate, as will be shown in part IV, along with the reorientational barrier $E_{BII}$, viz. the energy difference between a maximum and a minimum on $E_L$ (respectively $E_U$),

$$E_{BII} \equiv E_{Lmax} - E_{Lmin} \equiv E_{Umax} - E_{umin} = (I\omega^2/2)[(d_b - d_c)/b] Q_A^2 = (I\omega_{renII}^2/8) \qquad (8)$$

We also define the crossover energy

$$E_{CII} = E_{BII} + \tfrac{1}{2} E_{12} = [(d_b + d_c)/(d_b - d_c)] E_{BII} + 2E_{JT} \qquad (9)$$

The optical excitation energy of an off-center impurity in ground vibronic state of $E_L$ to the top of $E_U$

$$E_{OII} \equiv E_{Umax} - E_{Lmin} = 2E_{BII} + E_{12} = (I\omega_{renII}^2/4) + 2\{(I\omega_{renII}^2/4)[d_c/(d_b - d_c)] + 2E_{JT}\} \qquad (10)$$

exceeds $4E_{JT} = 2b^2/M\omega_{bare}^2$, the optical energy of the off-center effect by first-order electron-phonon coupling [2]. Both $E_{12}$ and $E_{OII}$ are composed of rotational parts proportional to I and vibrational parts to $E_{JT}$, while $E_{BII}$ is purely rotational in nature, as it should. The azimuthal dependence of the two-branch vibronic potential energy in equation (1)

$$E_\pm(\varphi) = \pm (I_A \omega_A^2 / b) \{-(d_b - d_c)[\tfrac{1}{4}(3 + \cos(4\varphi))] + d_b\} Q_A^2 +$$

$$\tfrac{1}{2}[(1\pm 2)(b^2/M_A \omega_A^2) - M_A \omega_A^2 E_{\alpha\beta}^2/4b^2] \qquad (11)$$

is depicted in Figure 1 marking the quantities introduced above.

## 2. 2. Solutions to the rotational equation

### 2. 2.1. General solution

We next seek solutions to the (1)-based Schrödinger equation. Rewriting to sort out the constant terms we get

$$- ( \eta^2 / 2I_A ) ( \partial^2 u / \partial \varphi^2 ) + 2B_\pm \cos(4\varphi)u + (C_\pm - E)u = 0 \qquad (12)$$

$$B_\pm = \mu (I_A \omega_A^2 / 8) [ ( d_b - d_c ) / b]Q_A^2 = \mu (E_{BII} / 4)$$

$$C_\pm = \pm \tfrac{1}{2} E_{BII} [ ( 3d_c + d_b ) / ( d_b - d_c )] + E_{JT} [(1\pm 2) - ( E_{\alpha\beta} / 4E_{JT} )^2 ]$$

where $u = u(\varphi)$ is the solution. Eq.(11) is Mathieu's equation and its solutions are known as Mathieu functions [8-11] (see Appendix).

At $B \propto ( d_b - d_c ) \equiv 0$ the eigenspectrum is one of a free rigid 2D rotator (cf. eq.(I.18))

$$E_n = (\eta^2 n^2 / 2I_A ) + C_\pm \qquad (13)$$

while the general periodic solution to equation (11) reads

$$u_n (\varphi,0) = A_c \cos(n\varphi) + A_s \sin(n\varphi) \qquad (14)$$

where n is any integer.

In conventional notations Mathieu's equation reads [9]:

$$d^2 Y / dz^2 + (a - 2q \cos 2z)Y = 0,$$

the relationship between different notations being $z = 2\varphi$, $Y = u$,

$$q = 2 ( B_\pm I_A / \eta^2 ) = \mu ( 2E_B / \eta\omega_{renII} )^2$$

$$\{\alpha\}_m = 2 ( I_A / \eta^2 ) (E_\alpha - C_\pm)$$

$$\{\beta\}_m = 2 (I_A / 2\eta^2) (E_\beta - C_\pm) \qquad (15)$$

There are two types of periodic solutions, even $ce_m (z,q)$ and odd $se_m (z,q)$ with eigenvalues $a_m (q)$ and $b_m (q)$, respectively. These functions will be normalized so that:

$$(1/\pi) \int_0^{2\pi} Y_m^2 (x) dx = 1$$

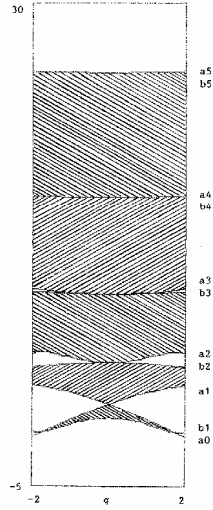

Figure 2. Allowed rotational energy bands obtained for q < 0 (upper sign branch) and q > 0 (lower sign branch), as calculated by means of Mathieu's eigenvalue expansions in Appendix.

At small q ≪ 1, $ce_0(z,q) \approx 1$, $ce_m(z,q) \approx \cos(mz)$, $se_m(z,q) \approx \sin(mz)$, and $\alpha_m \approx m^2$, $\beta_m \approx m^2$. These features help to identify m with the quantum number n in equation (13). We get two sets of eigenvalue spectra reading

$$E_{am} = (\eta^2 / 2I_A) a_m + C_\pm$$

$$E_{bm} = (\eta^2 / 2I_A) b_m + C_\pm, \quad (16)$$

respectively, where $a_m$, $b_m$ are generally dependent on q. The eigenvalue expansions in q are known at both small and large q (see Appendix). We remind that

$$q = (I_A \omega_{renII}^2 / 4 \eta \omega_{renII})^2 = (2E_{BII} / \eta \omega_{renII})^2$$

is proportional to the squared ratio of the renormalized rotational and vibrational energies. q = 4 is the critical value at which $E_{BII} = \eta \omega_{renII}$.

Mathieu's eigenvalue equations (11) define allowed energy bands corresponding to q < 0 (upper-sign branch) and q > 0 (lower sign branch), respectively [3]. These bands display the general features known from textbooks on band theory, in particular they widen as m is increased at fixed q. As $q = (2E_{BII} / \eta \omega_{renII})^2$ is increased, however, the allowed bands get narrower tending to turn into single levels in the limit of too high barriers. The borderlines of the allowed regions in the energy-versus-q plane are described by Mathieu's parameters $a_m(q)$ or $b_m(q)$, or, alternatively, by their corresponding periodic eigenfunctions $ce_m(z,q)$ or $se_m(z,q)$ at consecutive integral m (m = 1, 2, 3, 4,...), while each respective line in the interior is generated by pairs of Mathieu functions $ce_m(z,q)$ and $se_m(z,q)$ at an intermediate nonintegral m. Mathieu's eigenfunctions $ce_m(z,q)$ and $se_m(z,q)$ each have m zeros in the $0 < z < S\pi$ interval, and either a basic function or its derivative is vanishing at z = S π [9]. Within each allowed energy band, Mathieu's eigenstates and eigenvalues are functions of the wavenumber k = mπ where m runs from 0 to 1 for the first Brillouin zone [3].

We distinguish between two cases corresponding to the upper- and lower-sign branches in equation (11), respectively: In the former "negative q" case, the allowed bands are borderlined as follows: $(a_0,a_1)$, $(b_1,b_2)$, $(a_2,a_3)$, $(b_3,b_4)$, etc., while in the latter "positive q" case the border pairs are $(a_0,b_1)$, $(a_1,b_2)$, $(a_2,b_3)$, $(a_3,b_4)$, etc. When q is large each pair of borderlines approach one another and we can attach to the bands e.g. linear combinations of borderline eigenstates of the form, respectively: ½ ($ce_0 + ce_1$), ½ ($se_1 + se_2$), ½ ($ce_2 + ce_3$), ½ ($se_3 + se_4$), etc. for the "negative q", and of the form: ½ ($ce_0 + se_1$), ½ ($ce_1 + se_2$), ½ ($ce_2 + se_3$), ½ ($ce_3 + se_4$), etc. for the "positive q". Each linear combination is again an eigenstate whose energy falls within a "squeezed band". It is obvious that the allowed bands in the "negative q" case are of definite alternating parities, even, odd, etc. in the increasing m order, while the "positive q" bands are mixed parity. Examples of allowed rotational bands as calculated using the series' expansions in Appendix are presented in Figure 2.

### 2.2.2. Solution at the well bottom

The vibronic equation

$$- (\eta^2 / 2I_A) (\partial^2 u / \partial \varphi^2) \mu (I_A \omega_{renII}^2 / 16) \cos(4\varphi) u + \{ \pm (I_A \omega_{renII}^2 / 16) [(3d_c + d_b) / (d_b - d_c)] +$$
$$½ [(1 \pm 2)(b^2 / M\omega^2) - M\omega^2 E_{\alpha\beta}^2 / 4b^2] \} u = Eu \quad (17)$$

converts to a harmonic-oscillator type at small $|\varphi| < \pi /3.6$:

$$- (\eta^2 / 2I_A) (\partial^2 u / \partial \varphi^2) \mu (I_A \omega_{renII}^2 / 2) (¼ \pi - \varphi)^2 u + \{ \pm (I_A \omega_{renII}^2 / 8) [(d_b + d_c) / (d_b - d_c)] +$$
$$½ [(1 \pm 2)(b^2 / M\omega^2) - M\omega^2 E_{\alpha\beta}^2 / 4b^2] \} u = Eu \quad (18)$$

having introduced $\varphi ' = ¼ \pi - \varphi$ to get $\cos(4\varphi') \sim 1 - (1/2!)(4\varphi')^2$ around a well bottom at $\varphi = ¼ \pi$. Equation (18), e.g. for the lower-sign branch, is one of a harmonic oscillator. Defining a dimensionless coordinate

$$X = (K_{renII} Q_A^2 / \eta \omega_{renII})^{½} \varphi = (I_A \omega_{renII} / \eta)^{1/2} \varphi, \quad (19)$$

this equation turns into

$$- (\eta \omega_{renII} / 2) (\partial^2 u / \partial X^2) + (\eta \omega_{renII} / 2) [X - (I_A \omega_{renII} / \eta)^{1/2} (¼ \pi)]^2 u -$$
$$\{ (I_A \omega_{renII}^2 / 8) (d_b + d_c) / (d_b - d_c) + ½ [(b^2 / M\omega^2) + M\omega^2 E_{\alpha\beta}^2 / 4b^2] \} u = Eu \quad (20)$$

with eigenvalues

$$E_n = (n + ½) \eta \omega_{renII} - \{(I_A \omega_{renII}^2 / 8) [(d_b + d_c)/(d_b - d_c)] + ½[(b^2/M\omega^2) + M\omega^2 E_{\alpha\beta}^2 / 4b^2]\} \quad (21)$$

and eigenstates

$$u_n (\varphi) = N_n H_n(X) \exp(-X^2 / 2) = [(I_A \omega_{renII} / \eta)^{1/2} / \pi^{1/2} 2^n n!]^{½} H_n(X) \exp(-X^2 / 2) \quad (22)$$

where $H_n(X)$ stands for the Hermite polynomial of n-th order.

For calculating a relaxation rate, a few other quantities in the harmonic approximation are of importance: the lattice relaxation energy defined as the energy expended on creating two neighboring reorientational sites, e.g. at $\varphi_1 = ¼ \pi$ and $\varphi_2 = -¼ \pi$.

$$E_{RII} \equiv E_L(\varphi = -¼ \pi) - E_L(\varphi = ¼ \pi) = \pi^2 E_{BII}, \tag{23}$$

with

$$E_L(\varphi) = (I_A \omega_{ren}^2/2)(¼ \pi - \varphi)^2 u - \{(I_A \omega_{renII}^2 / 8)[(d_b + d_c)/(d_b - d_c)] +$$

$$½ [(b^2 / M\omega^2) + M\omega^2 E_{\alpha\beta}^2 / 4b^2]\}$$

$E_R$ is only meaningful at a well bottom where the potential energy is nearly parabolic with a renormalized curvature.

From (I.24) and (I.8) we have

$$E_{BII} / [ ½ \eta\omega_{renII}] = [½ KQ_A^2 / \eta\omega_{bare}][(d_b - d_c)Q_A^3 / bQ_A]^{1/2}$$

which implies that a certain part of the elastic energy should exceed the bare phonon quantum if the vibronic system is to have underbarrier energy levels at $E_{BII} > ½ \eta\omega_{renII}$.

Calculated rotational parameters associated with off-center ions in fcc alkali halides are presented in Table I (isolated impurity) and Table II (impurity at F center).

### 2.3. Electrostatic polarizability of off-center impurity

The splitting of a normal lattice site into off-center sites to be occupied by a small-radius cation impurity gives rise to an off-center ellipsoid. As a matter of fact, on minimizing

$$E_L(\{Q_i\}) = ½ \{ \sum_i K_I Q_i^2 - [ \sum_i (2b_iQ_i)^2 + E_{\alpha\beta}^2 ]^{½} \} \tag{24}$$

with respect to $Q_k$ we obtain

$$Q_x^2 / Q_{x0}^2 + Q_y^2 / Q_{y0}^2 + Q_z^2 / Q_{z0}^2 = 1 \tag{25}$$

with semiaxes

$$Q_{x0} = [(2E_{JTx} / K_x)(1 - \mu_x^2)]^{1/2}, \mu_x = E_{\alpha\beta} / 4E_{JTx}$$

$$Q_{y0} = [(2E_{JTy} / K_y)(1 - \mu_y^2)]^{1/2}, \mu_y = E_{\alpha\beta} / 4E_{JTy}$$

$$Q_{z0} = [(2E_{JTz} / K_z)(1 - \mu_z^2)]^{1/2}, \mu_z = E_{\alpha\beta} / 4E_{JTz} \tag{26}$$

This ellipsoid is the locus of stable spatial impurity positions off-centered relative to the normal lattice site. An off-center site instability clearly occurs at $\mu_x, \mu_y, \mu_z \leq 1$ where $E_{JTi} = b_i^2/2K_i$ are

Jahn-Teller (JT) energies, $K_i = M\omega_i^2$, and $\omega_i$ are the bare-phonon frequencies associated with $Q_i$.

An off-centered ellipsoid is polarizable electrostatically. Its coupling energy to an electric field will be computed as

$$U_C = \tfrac{1}{2}\, \hat{\alpha}\, \mathbf{F}.\mathbf{F} = \tfrac{1}{2} \sum_{ij} \alpha_{ij} F_i F_j \qquad (27)$$

where $\hat{\alpha}$ is the polarizability tensor of the off-center ellipsoid, while $\mathbf{F}$ is the electric field.

The off-centered ion traverses across the off-center volume all around the normal lattice site which modifies its polarizability. Consequently $\hat{\alpha}$ is the polarizability tensor of an off-center ellipsoid rather than the one of a single ion. To deduce its components, we assume coupling to the $Q_i$ modes. For instance, coupling to the ungerade bending modes $Q_x$ and $Q_y$ of the <110> $Cl^-$-$Li^+$-$Cl^-$ bonds will yield four off-axis sites in the basal (x,y)-plane of an $F_A$ center, with [5]:

$$\alpha_{xy} = [P_E^2(1-\mu_E^2)/3t_E]\{(t_E/k_B T)\exp(-t_E/k_B T) + \sinh(t_E/k_B T)\}/\{\exp(-t_E/k_B T) + \cosh(t_E/k_B T)\} \qquad (28)$$

defined by means of the mixing dipoles

$$P_E \equiv P_{x(y)}\, \delta n_{x(y)} = <a_{1g} | ex(y) | t_{1ux(y)}> \delta n_{x(y)}$$

and the tunneling splitting $t_E$. $t_E$ is associated with transfers between off-center sites and is defined by the respective JT energies, spring- and mixing- constants [14]

$$t_E = E_{JTE}\, \mu_E\, (1 - \mu_E) / \sinh(u_E^2)$$

$$u_E = [(2E_{JTE} / \eta\omega_E)(1 - \mu_E^2)^{1/2}]^{1/2} \qquad (29)$$

$\delta n_{x(y)} = n_{t1ux(y)} - n_{a1g}$ are the differences in occupation numbers between $t_{1ux(y)}$ and $a_{1g}$.

$\alpha_{xy}$ is the electrostatic polarizability of the off-centered Li-impurity in (x,y)-plane underlying an F center in [100] site, giving rise to an off-centered polarizable circle. Therefore it may be essential in experiments on the electrostatic response of Li-impurities and $F_A(Li)$ centers in alkali halides, e.g. by paraelectric resonance [15]. It has a specific temperature dependence with a lower-temperature portion at $k_B T \ll t_E$:

$$\alpha_{xy}(0) = P_E^2 (1 - \mu_E^2)/3t_E \qquad (30)$$

and a higher-temperature branch at $k_B T \gg t_E$:

$$\alpha_{xy}(T) = P_E^2 (1 - \mu_E^2) / k_B T, \qquad (31)$$

as shown elsewhere [5].

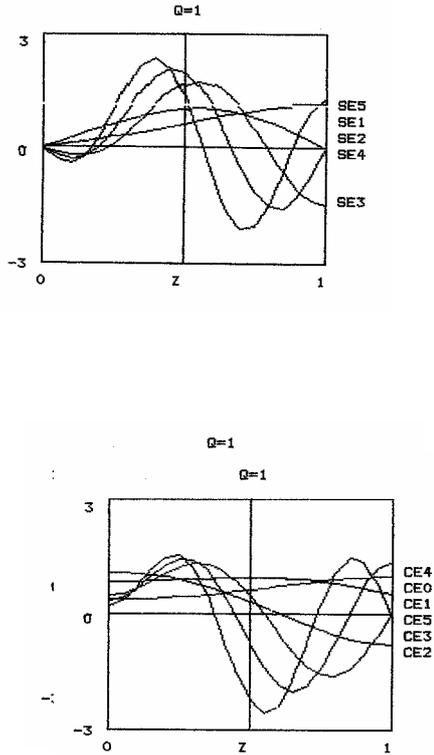

Figure 3. Calculated examples of azymuthal φ-dependences of Mathieu's periodic eigenfunctions, odd parity $se_m(z,q)$ and even parity $ce_m(z,q)$ using the series expansions at "not too large q", as given in Appendix. The abscissa is $Z = (2/\pi)z = (4/\pi)\varphi$.

Appendix

Mathieu functions

There are two types of periodic solutions to Mathieu's equation, even $ce_m(z,q)$ and odd $se_m(z,q)$, with eigenvalues $a_m(q)$ and $b_m(q)$, respectively [9]. Alternative notations are also used as follows: $ce_{2n}(z,q)$, $se_{2n+1}(z,q)$, $ce_{2n+1}(z,q)$, and $se_{2n+2}(z,q)$ for n = 0,1,2,... with eigenvalues $\alpha_{2n}$, $\beta_{2n+1}$, $\alpha_{2n+1}$, and $\beta_{2n+2}$, respectively ($q = 8q$, $a_r = 4\alpha_r$, $b_r = 4\beta_r$) [10]. Graphic examples of periodic Mathieu functions at "not too large q" are shown in Figure 3.

These functions are often normalized so that:

$(2/\pi) \int_0^{\pi/2} Y_m^2(x)dx = 1$ (m=0), $= \frac{1}{2}$ (m=1,2,3,…) [10]

$(1/\pi) \int_0^{2\pi} Y_m^2(x)dx = 1$ [11]

At small $q \ll 1$, $ce_0(z,q) \sim 1$, $ce_m(z,q) \sim \cos(mz)$, $se_m(z,q) \sim \sin(mz)$, and $4\alpha_m \sim m^2$, $\beta_m \sim m^2$. Eigenvalue expansions in q are available at both small and large q. For instance, the first few eigenvalues expand at "not too large q" such that:

$a_0(q) = -\frac{1}{2}q^2 + (7/128)q^4 - (29/2304)q^6 + (68687/18874368)q^8 +...$

$a_1(-q) \equiv b_1(q) = 1 - q - (1/8)q^2 + (1/64)q^3 - (1/1536)q^4 - (11/36864)q^5 + (49/589824)q^6 -$

$(55/9437184)q^7 - (83/35389440)q^8 +...$

$b_2(q) = 4 - (1/12)q^2 + (5/13824)q^4 - (289/79626240)q^6 + (21391/458647142400)q^8 +...$

$a_2(q) = 4 + (5/12)q^2 - (763/13824)q^4 + (1002401/79626240)q^6 - (1669068401/458647142400)q^8 +...$

$a_3(-q) \equiv b_3(q) = 9 + (1/16)q^2 - (1/64)q^3 + (13/20480)q^4 + (5/16384)q^5 - (1961/23592960)q^6 + (609/104857600)q^7 +...$

$b_4(q) = 16 + (1/30)q^2 - (317/864000)q^4 + (10049/2721600000)q^6 +...$

$a_4(q) = 16 + (1/30)q^2 + (433/864000)q^4 - (5701/2721600000)q^6 +...$

$a_5(-q) \equiv b_5(q) = 25 + (1/48)q^2 + (11/774144)q^4 - (1/147456)q^5 + (37/891813888)q^6 +...$

$b_6(q) = 36 + (1/70)q^2 + (187/43904000)q^4 - (5861633/92935987200000)q^6 +...$

$a_6(q) = 36 + (1/70)q^2 + (187/43904000)q^4 + (6743617/92935987200000)q^8 +...$

$a_r(q) \equiv b_r(q) = r^2 + [1/2(r^2-1)]q^2 + [(5r^2+7)/32(r^2-1)^3(r^2-4)]q^4 + [(9r^4+58r^2+29)/64(r^2-1)^5(r^2-4)(r^2-9)]q^6 +...$ $(r \geq 7)$.

The Mathieu functions expand in z at not too large q as [9]:

$ce_0(z,q) = 2^{-1/2}\{1 - (1/2)q\cos(2z) + q^2[\cos(4z)/32 - 1/16] - q^3[\cos(6z)/1152 - 11\cos(2z)/128] +...\}$

$ce_1(z,q) = \cos(z) - (1/8)q\cos(3z) + q^2[\cos(5z)/192 - \cos(3z)/64 - \cos(z)/128] - q^3[\cos(7z)/9216 - \cos(5z)/1152 - \cos(3z)/3072 + \cos(z)/512] +...$

$se_1(z,q) = \sin(z) - (1/8)q\sin(3z) + q^2[\sin(5z)/192 + \sin(3z)/64 - \sin(z)/128] - q^3[\sin(7z)/9216 + \sin(5z)/1152 - \sin(3z)/3072 - \sin(z)/512] +...$

$ce_2(z,q) = \cos(2z) - q[\cos(4z)/12 - 1/4] + q^2[\cos(6z)/384 - 19\cos(2z)/288] +...$

$se_2(z,q) = \sin(2z) - (1/12)q\sin(4z) + q^2[\sin(6z)/384 - \sin(2z)/288] +...$

$ce_r(z,q)\ (p=0) \equiv se_r(z,q)\ (p=1) = \cos(rz-p\pi/2) - q\{\cos[(r+2)z-p\pi/2]/4(r+1) - \cos[(r-2)z-p\pi/2]/4(r-1)\} + q^2\{\cos[(r+4)z-p\pi/2]/32(r+1)(r+2) + \cos[(r-4)z-p\pi/2]/32(r-1)(r-2) - [\cos(rz-p\pi/2)/32][2(r^2+1)/(r^2-1)^2]\} +...$ $(r \geq 3)$.

Anticipating further needs we list the respective derivatives:

$dce_0(z,q)/dz = 2^{-1/2}\{q\sin(2z) - (1/8)q^2\sin(4z) - q^3[-(6/1152)\sin(6z) + (22/128)\sin(2z)]\}$

$dce_1(z,q)/dz = -\sin(z) + (3/8)q\sin(3z) + q^2[-(5/192)\sin(5z) + (3/64)\sin(3z) + (1/128)\sin(z)] - q^3[-(7/9216)\sin(7z) + (5/1152)\sin(5z) + (3/3072)\sin(3z) - (1/512)\sin(z)]$

$dce_2(z,q)/dz = -2\sin(2z) + (1/3)q\sin(4z) + q^2[-(6/384)\sin(6z) + (38/288)\sin(2z)]$

$dce_r(z,q)/dz = -r\sin(rz) - (1/4)q\{-[(r+2)/(r+1)]\sin[(r+2)z] + [(r-2)/(r-1)]\sin[(r-2)z]\} + (1/32)q^2\{-[(r+4)/(r+1)(r+2)]\sin[(r+4)z] - [(r-4)/(r-1)(r-2)]\sin[(r-4)z] + [2r(r^2+1)/(r^2-1)^2]\sin(rz)\}$

$dse_1(z,q)/dz = \cos(z) - (3/8)q\cos(3z) + q^2[(5/192)\cos(5z) + (3/64)\cos(3z) - (1/128)\cos(z)] - q^3[(7/9216)\cos(7z) + (5/1152)\cos(5z) - (3/3072)\cos(3z) - (1/512)\cos(z)]$

$dse_2(z,q)/dz = 2\cos(2z) - (1/3)q\cos(4z) + q^2[(6/384)\cos(6z) - (2/288)\cos(2z)]$

$dse_r(z,q)/dz = -r\sin(rz-\pi/2) - (1/4)q\{-[(r+2)/(r+1)]\sin[(r+2)z-\pi/2] + [(r-2)/(r-1)]\sin[(r-2)z-\pi/2]\} + (1/32)q^2\{-[(r+4)/(r+1)(r+2)]\sin[(r+4)z-\pi/2] - [(r-4)/(r-1)(r-2)]\sin[(r-4)z-\pi/2] + [2r(r^2+1)/(r^2-1)^2]\sin[rz-\pi/2]\}$ for $r \geq 3$.

Using the above expansions we calculate finite-valued saddle-point functions at $z = \pm S\pi$ ($\varphi = \pm j\pi$), even

$ce_0(\pm S\pi,q) = 2^{-1/2}[1 + (1/2)q - (1/32)q^2 - 0.085069(4)q^3]$

$ce_2(\pm S\pi,q) = -1 + (1/6)q + 0.0633680(5)q^2$

$ce_r(\pm S\pi,q) = (-1)^{r/2} - (1/4)q\{(-1)^{r/2}+1/(r+1) - (-1)^{r/2}-1/(r-1)\} + (1/32)q^2\{(-1)^{r/2}+2/(r+1)(r+2) + (-1)^{r/2}-2/(r-1)(r-2) - (-1)^{r/2}[2(r^2+1)/(r^2-1)^2]\}$, $r = 2n$

and odd

$se_1(\pm\pi/2,q) = \pm[1 + (1/8)q - 0.0182291(6)q^2 - 6.51041(6)\times10^{-4}q^3]$

$se_r(\pm\pi/2,q) = \pm\{(-1)^{(r-1)/2} - (1/4)q[(-1)^{(r+1)/2}/(r+1) - (-1)^{(r-3)/2}/(r-1)] + (1/32)q^2\{(-1)^{(r+3)/2}/(r+1)(r+2) + (-1)^{(r-5)/2}/(r-1)(r-2) - (-1)^{(r-1)/2}[2(r^2+1)/(r^2-1)^2]\}\}$, $r = 2n-1$,

as well as derivatives, odd

$dce_1(\pm\pi/2,q)/dz = \pm[-1 - (3/8)q - 0.0651041(6)q^2 - 2.17013(8)\times10^{-3}q^3]$

$dce_r(\pm\pi/2,q)/dz = \pm\{-r(-1)^{(r-1)/2} - jq[-(-1)^{(r+1)/2}(r+2)/(r+1) + (-1)^{(r-3)/2}(r-2)/(r-1)] +$

$(1/32)q^2\{-(-1)^{(r+3)/2} (r+4)/(r+1)(r+2) - (-1)^{(r-5)/2} (r-4)/(r-1)(r-2) + (-1)^{(r-1)/2}[2r(r^2+1)/(r^2-1)^2]\}\}, r=2n-1$

and even

$dse_2(\pm \pi/2, q)/dz = -2 - (1/3)q - 8.680(5) \times 10^{-3} q^2$

$dse_r(\pm \pi/2, q)/dz = -r(-1)^{r/2-1} - jq\{-(-1)^{r/2}(r+2)/(r+1) + (-1)^{r/2+2}(r-2)/(r-1)\} +$

$(1/32)q^2\{(-1)^{r/2+1}(r+4)/(r+1)(r+2) - (-1)^{r/2-3}(r-4)/(r-1)(r-2) + (-1)^{r/2-1} 2r(r^2+1)/(r^2-1)^2\}, r = 2n$

The remaining saddle-point values are all vanishing, namely,

$ce_1(\pm \pi/2, q) = 0$

$ce_r(\pm \pi/2, q) = 0, r = 2n-1$

$se_2(\pm \pi/2, q) = 0$

$se_r(\pm \pi/2, q) = 0, r = 2n$

$dce_0(\pm \pi/2, q)/dz = 0$

$dce_2(\pm \pi/2, q)/dz = 0$

$dce_r(\pm \pi/2, q)/dz = 0, r = 2n$

$dse_1(\pm \pi/2, q)/dz = 0$

$dse_r(\pm \pi/2, q)/dz = 0, r = 2n-1$

Finally, the relationships between Mathieu's functions at negative and positive q read:

$ce_{2n}(z, -q) = (-1)^n ce_{2n}(S \pi - z, q)$

$ce_{2n+1}(z, -q) = (-1)^n se_{2n+1}(S \pi - z, q)$

$se_{2n}(z, -q) = (-1)^n se_{2n}(S \pi - z, q)$

$se_{2n+1}(z, -q) = (-1)^n ce_{2n+1}(S \pi - z, q).$

Other graphic examples of periodic Mathieu functions can be found in textbooks on transcendent functions [9,10].

## References

[1] G. Baldacchini, R.M. Montereali, U.M. Grassano, A. Scacco, P. Petrova, M. Mladenova, M. Ivanovich, and M. Georgiev, cond-mat 0709.1951 (preceding Part I).

[2] G. Baldacchini, U.M. Grassano, A. Scacco, F. Somma, M. Staikova, M. Georgiev, Nuovo Cim.**13 D** 1399-1421 (1991).
[3] L. Brillouin} and M. Parodi, Propagation des Ondes dans les Milieux Periodiques (Dunod, Paris, 1956). Russian translation: L. Brillyuen} i M. Parodi, Rasprostranenie voln v Periodicheskikh strukturakh (IIL, Moskva, 1959, p.p. 266-306).
[4] M. Gomez, S.P. Bowen, J.A. Krumhansl, Phys. Rev. **153** 1009 (1967).
[5] I.B. Bersuker, The Jahn-Teller Effect and Vibronic Interactions in Modern Chemistry (Academic, New York, 1986). Russian translation: Effekt Yana-Tellera i Vibronnie Vzaimodeystviya v Sovremennoi Khimii (Nauka, Moskva, 1988).
[6] L. Pauling} and E. Bright Wilson, Introduction to Quantum Mechanics (Dover, New York, 1985), p.p. 291-292.
[7] H. Eyring, J. Walter, and G.E. Kimball, Quantum Chemistry (Wiley, New York, 1944), p.p. 358-360.
[8] E. Kamke, Gewöhnliche Differentialgleichungen (Leipzig, 1959). Russian translation: (Nauka, Moscow, 1971).
[9] M. Abramowitz & I.A. Stegun, eds. Handbook of Mathematical Functions with Formulas, Graphs and Mathematical Tables}(NBS Math. Series, 1964). Russian translation: (Nauka, Moscow, 1979).
[10] E. Janke, F. Emde, F. Loesch, Tafeln Höherer Funktionen (Teubner, Stuttgart, 1960). Russian translation: (Nauka, Moscow, 1964).
[11] I.S. Gradsteyn and I.M. Ryzhik, Tablitsy integralov, summ, ryadov i proizvedenii}. (GIFML, Moskva, 1963). English: I.S. Gradshteyn and I.M. Ryzhik, Table of Integrals, Series, and Products, Corrected and Enlarged Edition (Academic Press, Orlando, 1965).
[12] L.S. Bark, N.I. Dmitrieva, L.N. Zakhar'ev, A.A. Lemanskii, Tablitsi sobstvennykh znachenii uravneniya Mat'e} (Computing Center of the USSR Academy of Sciences, Moscow, 1970).
[13] M.J.O. Strutt, Lamesche-Mathieusche und Verwandte Funktionen in Physik und Technik (Russian Translation: GNTIU, Kharkov-Kiev, 1935).
[14] L. Mihailov, M. Ivanovitch, and M. Georgiev, Physica C **223**, 249-258 (1994).
[15] F. Lüty, J. Physique (Paris) **28,** C4-120 (1967).

Table I

Calculated rotational parameters

Isolated impurity

| Host | Off-center radius $Q_0$ (Å) | II renorm. frequency $\omega_{renII}$ ($10^{13}$ s$^{-1}$) | Inertial moment $I \times 10^{26}$ (eV×m$^2$) | Rotatiion barrier $E_{BII}$ (eV) | Adiabatic energy splitting $E_{12}$ (eV) | Crossover energy $E_{CII}$ (eV) | Optical energy $E_{0II}$ (eV) |
|---|---|---|---|---|---|---|---|
| LiF | 0.7660 | 6.7756 | 0.1743 | 1.0002 | 10.5556 | 6.2780 | 12.5561 |
| NaF | 0.6774 | 4.1810 | 0.1363 | 0.2978 | 5.8884 | 3.2420 | 6.4841 |
| KF | 0.7717 | 3.1843 | 0.1769 | 0.2242 | 5.8001 | 3.1242 | 5.1986 |
| RbF | 1.0510 | 3.3831 | 0.3281 | 0.4694 | 5.4801 | 3.2094 | 6.4190 |
| LiCl | 0.3222 | 1.3949 | 0.0575 | 0.0140 | 7.5521 | 3.7900 | 7.5801 |
| " | 0.7113 | 3.0795 | 0.2804 | 0.3324 | 7.1365 | 3.9006 | 7.8013 |

| Host | | | | | | | |
|---|---|---|---|---|---|---|---|
| NaCl | 0.5508 | 1.9265 | 0.1279 | 0.0593 | 3.8673 | 1.9930 | 3.9859 |
| KCl | 0.4096 | 1.1273 | 0.0930 | 0.0148 | 2.3011 | 1.1653 | 2.3307 |
| RbCl | 0.5162 | 1.1554 | 0.1477 | 0.0246 | 2.4510 | 1.2501 | 2.5003 |
| NaBr | 0.3279 | 0.8622 | 0.1343 | 0.0125 | 2.3713 | 1.1981 | 2.3963 |
| KBr | 0.1981 | 0.4018 | 0.0490 | 0.0010 | 1.4879 | 0.7449 | 1.4899 |
| RbBr | 0.2894 | 0.4713 | 0.1046 | 0.0029 | 1.6274 | 0.8166 | 1.6332 |
| NaI | 0.1472 | 0.3084 | 0.0430 | 0.0005 | 1.6021 | 0.8015 | 1.6032 |

Table II

Calculated rotational parameters

Impurity at F center

| Host | Off-center radius $Q_A$[a] (Å) | II renorm. frequency $\omega_{renIIA}$[b] ($10^{13}$ s$^{-1}$) | Inertial moment $I_A*10^{26}$[c] (eV*m$^2$) | Rotation barrier $E_{BIIA}$[d] (eV) | Adiabatic energy splitting $E_{12A}$[e] (eV) | Crossover energy $E_{CIIA}$[f] (eV) | Optic energy $E_{OII}$[g] (eV) |
|---|---|---|---|---|---|---|---|
| LiF | 0.7548 | 8.1770 | 0.1128 | 0.9428 | 10.6306 | 6.2581 | 12.5162 |
| NaF | 0.6634 | 5.0148 | 0.0871 | 0.2738 | 5.9198 | 3.2337 | 6.4674 |
| KF | 0.7563 | 3.8221 | 0.1133 | 0.2069 | 4.7727 | 2.5933 | 5.1865 |
| RbF | 1.0404 | 4.1017 | 0.2143 | 0.4507 | 5.5046 | 3.2030 | 6.4060 |
| LiCl | 0.6917 | 3.6676 | 0.1768 | 0.2973 | 7.1823 | 3.8885 | 7.7769 |
| NaCl | 0.5286 | 2.2643 | 0.0785 | 0.0503 | 3.8791 | 1.9899 | 3.9797 |
| KCl | 0.3703 | 1.2481 | 0.0507 | 0.0099 | 2.3075 | 1.1637 | 2.3273 |
| RbCl | 0.4862 | 1.3328 | 0.0873 | .0194 | 2.4579 | 1.2484 | 2.4967 |
| NaBr | 0.2774 | 0.8933 | 0.0641 | 0.0064 | 2.3792 | 1.1960 | 2.3920 |
| KBr | 0.1034 | 0.2568 | 0.0089 | $7\times10^{-5}$ | 1.4891 | 0.7446 | 1.4892 |
| RbBr | 0.2302 | 0.4592 | 0.0441 | 0.0012 | 1.6296 | 0.8160 | 1.6320 |

[a] $Q_A = Q_0 \{1 - (\alpha_F p_{sp}^2/E_{JT}k^2R^6)/(4E_{JT}/E_{\alpha\beta})^2 -1]\}$: $\alpha_F$, $p_{sp}\equiv p_z$ data from Ref. [2], $Q_0$, $E_{JT}$, $E_{\alpha\beta}$, k=$k_p$, R≡$r_0$ data from Ref. [1]. [b] $\omega_{renIIA} = 1.5\omega_{renII}$ ($Q_A/Q_0$), the 1.5 factor arising from the dimensionality reduction from 3D to 2D [2]. [c] $I_A = M_A Q_A^2 = (2/3)I(Q_A/Q_0)^2$. [d] $E_{BIIA} = I_A \omega_{renIIA}^2 / 8$. [e] $E_{12} = 4[E_{BIIA} d_c / (d_b-d_c) + E_{JT}]$. [f] $E_{CIIA} = E_{BIIA} + (1/2)E_{12A}$. [g] $E_{OIIA} = 2E_{BIIA}+E_{12A}$.